\documentclass[aps, twocolumn,  showpacs]{revtex4}

\begin{document}
\title{ Weyl geometry, topology of space-time and reality of electromagnetic potentials, and new perspective on particle physics}
\author{S. C. Tiwari \\
Department of Physics, Institute of Science,  Banaras Hindu University,  and Institute of Natural Philosophy, \\
Varanasi 221005, India }
\begin{abstract}
Consistency of Weyl natural gauge, Lorentz gauge and nonlinear gauge is studied in Weyl geometry. Field equations in generalized Weyl-Dirac theory show that spinless electron and photon are topological defects. Statistical metric and fluctuating metric in 3D space with time as a measure of spatial relations are discussed to propose a statistical interpretation of Maxwell field equations. It is argued that physical geometry is an approximation to mathematical geometry, and 4D relativistic spacetime is essentially 3D space with changing spatial relations. The present work is suggested to have radical new outlook on elementary particle physics.  
\end{abstract}
\maketitle
 
 \section{\bf Introduction}
 
 Riemann, way back in 1861, had speculated on the physical significance of the electromagnetic (EM) potentials \cite{1}. Historical
 evolution of the concept of EM potentials has witnessed landmark achievements \cite{2, 3}, yet their physical reality remains unsettled. 
 Aharonov and Bohm in a comprehensive article \cite{4} present the role of geometry, topology, and measurability in connection with
 the physical reality of the potentials. Unfortunately the manner in which this question was introduced underlining classical versus
 quantum has become a source of confusion in the literature. Aharonov-Bohm (AB) effect has been demonstrated experimentally many times \cite{5},
 however multiple shades of philosophical/mathematical arguments tend to obscure the nature of the EM potentials \cite{6}.
 
 The issue has also acquired great significance in the context of the standard model (SM) of particle physics. A crucial element in SM
 and modern gauge theories is the postulated concept of hypothetical internal spaces for the gauge symmetry. For example, in SU(3) gauge group
 the set of parameters $\theta^a$ are Lorentz scalars and the index $a=1, 2, ...8$ for 8 generators of the group, and the unitary
 gauge transformation is $U_{SU(3)} = e^{i \theta^a (x)\lambda^a/2}$ where $\lambda^a$ are the Gell-Mann matrices. 
 In quantum electrodynamics (QED) there is just one
 parameter for the gauge transformation $U(1) = e^{i\theta (x)}$. If the parameters do not depend on space-time coordinate $x$ then we have
 global gauge transformation. Gauge invariance of the action function or the Lagrangian density leads to conserved Noether currents and
 corresponding gauge charges, for example, the electric charge is interpreted as a gauge charge for U(1) gauge symmetry in QED. 
 Note that QED is a paradigm for SM and
 modern gauge theories. However, originally the idea of gauge symmetry proposed by Weyl \cite{7} represented scale transformation of the line element
 in the pseudo-Riemannian spacetime of general relativity
 resulting into the change in the length of a vector under parallel transport. Thus gauge symmetry was a spacetime symmetry transformation with 
 a wider group of transformations than the general coordinate transformations of general relativity \cite{7}. We may now have two
 view-points. First, the empirical success of SM in precision experiments, and the detection of the gauge bosons of the electroweak unified theory
 could be taken to imply that
 the question of the reality of potentials has been rendered vacuous: the dynamical gauge potentials in this theory represent the gauge bosons i. e.
 photon and massive weak gauge bosons $W^\pm; Z$, and they have physical reality. This point is further elaborated in the next section.
 The second view could be as follows. The fundamental role of the internal spaces in modern theories, e. g.
 SM and superstrings, departs from the geometrization of physics based on the intuitive reality of space and time speculated by Riemann,
 Clifford, Einstein and Weyl \cite{1, 7}. Could one seek alternative scenario in that tradition? 
 
 The present work is focused on this question. Obviously the concepts of space and time would
 need radical revision in which nontrivial geometry and topology acquire central role \cite{8}. In the new perspective it emerges that classical Weyl geometry of space-time has not only a nontrivial topological structure it also admits statistical interpretation of geometric quantities. Appraisal of a comprehensive review on Weyl geometry \cite{9} and a critical reading of Dirac's paper \cite{10} show that most of the present work is a new contribution. The significance of our work lies in the fact that rather than discarding the reality of space and time we articulate a framework in which space and time constitute a fundamental physical reality.
 
 The paper is organized as follows. In the next section we explain the difference between geometry and topology; and discuss physical geometry for
 physical phenomena and its description. The conception of physical reality depends in a fundamental way on the approximation and/or limitation
 of the mathematical reality. The role of language in human affairs is primarily to express the thoughts and the feelings in a tangible
 form; this process has intrinsic limitation in the exact reproducibility of the subtle thoughts. The origin of thoughts, mind-brain
 duality, and the Platonic reality of mathematics invite attention to deep questions \cite{8}. Note that the symbolic logic had origin
 in Boole's work ``The Laws of Thought''. Our proposition that physics is the language of mathematics echoes the commonplace
 understanding regarding the expression of subtle thoughts in tangible form of the language. The philosophical discussions could be found in \cite{8}
 and references cited therein. Here the considerations are primarily concerned with electromagnetism and Weyl geometry leading to the important result that Weyl geometry could be endowed with a nontrivial topology and statistical nature.
 
 In section III we investigate various kinds of gauge conditions, i. e. Weyl's natural gauge, Lorentz gauge and nonlinear gauge, and their consistency in Weyl \cite{7} and Weyl-Dirac \cite{10} theories. The main result of this section is that the principle of gauge invariance may have statistical interpretation, and the elementary geometrical objects defined by pure gauge potentials may have a topology akin to that of AB effect \cite{4}. The field equations derived in a generalized Weyl-Dirac framework \cite{11} are discussed in section IV. Three important results are: electron could be decoupled from EM field and potentials, single electron could be visualized as a propagating topological defect, and new insight is gained on the topological model of photon \cite{12}.
 
 The physical significance of the nonlinear term in scalar curvature in Weyl geometry remains obscure \cite{7, 13}. In flat spacetime geometry Dirac in 1951 \cite{14} investigated the nonlinear gauge to develop a new theory of electron. On the other hand, Gubarev et al \cite{15} discussed topological structures for the minimum value of the volume integral of the squared potentials. The role of nonlinear gauge in the generalized Weyl-Dirac theory in connection with \cite{14, 15} is discussed in section V. A logically consistent approach is proposed for macroscopic system drawing analogy to fluid dynamics. In the last section a brief discussion on the implications of the present work on modern gauge field theories and the current ideas on emergent spacetime is presented followed by concluding remarks.
 
 \section{\bf Geometry and physics}

 Classical physics comprises of a classical system or object, observable/measurable physical quantities, and a theoretical description
 quite often having a mathematical formalism. A simple example is that of a macroscopic body that has well defined macroscopic variables
 determined by local differential equations. One may also consider a microscopic system and try to develop a macroscopic system from it
 using the methods of statistical mechanics. The intricate question is that of the necessary criteria characterizing a system 
 macroscopic or microscopic. 
 
 Majority of physicists follow Newton's experimental natural philosophy: experimentally measured quantities are defined in terms of mathematical
 variables and concepts, and depending on the physical laws or hypotheses the mathematical formalism is developed. In this method abstract
 mathematics is a convenient tool. It seems to have given rise to the belief that mathematics is a language of physics. In classical electrodynamics (CED) the 
 experimental force laws of Coulomb and Ampere, and Faraday's law of induction serve the basis for Maxwell equations
 \begin{equation}
  {\bf \nabla}.{\bf E} =\rho
 \end{equation}
 \begin{equation}
  {\bf \nabla} \times {\bf B} = {\bf J} + \frac{1}{c} \frac{\partial{\bf E}}{\partial t}
 \end{equation}
 \begin{equation}
  {\bf \nabla} \times {\bf E} = - \frac{1}{c} \frac{\partial{\bf B}}{\partial t}
 \end{equation}
 \begin{equation}
  {\bf \nabla}.{\bf B} =0 
 \end{equation}
 EM fields are mathematical variables defined from the force laws using a limiting procedure. First Chapter in the textbook \cite{16} rightly
 notes that independent physical reality of EM fields emerges when energy, momentum, and angular momentum associated with them are defined. We may
 add that only after the advent of special relativity and the observed mechanical effects of the radiation field separated from the sources
 $\rho$ and ${\bf J}$ the true physical significance of the EM fields was recognized and accepted. 

Mathematically, without recourse to CED, one may proceed defining a second-rank antisymmetric tensor $F_{\mu\nu}$ in 4D space
\begin{equation}
 F_{\mu\nu} = \partial_\mu A_\nu - \partial_\nu A_\mu
\end{equation}
and define a scalar
\begin{equation}
 \mathcal{L}= -\frac{1}{4} F_{\mu\nu} F^{\mu\nu}
\end{equation}
that gives the Euler-Lagrange equation from the action principle
\begin{equation}
 \partial_\mu F^{\mu\nu} =0
\end{equation}
Definition (5) gives
\begin{equation}
 \partial_\mu F_{\nu\sigma} + \partial_\nu F_{\sigma\mu} + \partial_\sigma F_{\mu\nu} =0
\end{equation}
The set of equations (5) to (8) constitutes a mathematical system. Formally they are equivalent to the source-free Maxwell equations
if 4D space is identified with spacetime: Eqs. (1) and (2) $\rightarrow$ Eq.(7); Eqs. (3) and (4) $\rightarrow$ Eq.(8). Physical
interpretation of the tensor $F_{\mu\nu}$ and the 4-vector $A_\mu$ representing EM phenomena is essential to identify Eqs. (7) and (8)
as Maxwell equations. Historically physics laws preceded Maxwell equations and it appeared as if the role of mathematics was just
that of a convenient description tool.

The constructs (5) and (6) look artificial because we already have the EM field tensor in mind, however Weyl theory \cite{7} presents
a new picture not recognized adequately. Spacetime metric geometry 
\begin{equation}
 ds^2 = g_{\mu\nu} dx^\mu dx^\nu
\end{equation}
is generalized postulating a linear groundform $W_\mu dx^\mu$ such that under gauge transformation
\begin{equation}
 ds \rightarrow ds^\prime =\lambda ~ ds
\end{equation}
\begin{equation}
 W_\mu \rightarrow W_\mu^\prime = W_\mu +\phi_{,\mu}
\end{equation}
where $\phi = \log \lambda$ and $\lambda$ is an arbitrary function of coordinates. Notation and metric convention are those of \cite{10, 13}. Note that in Weyl geometry the linear groundform is in addition to the metric groundform (9), and it depends on the gauge or calibration $\lambda$ via Eq.(11). In Weyl geometry both length and direction of a vector undergo changes under parallel transport from spacetime point $x^\mu$ to $x^\mu +dx^\mu$
and the total change in the length of a vector round a small closed loop is
\begin{equation}
 \oint \delta l = l W_{\mu\nu} \delta S^{\mu\nu}
\end{equation}
Here $\delta S^{\mu\nu}$ is the area enclosed by the loop and $W_{\mu\nu}$ is a geometric quantity termed distance curvature, see section 15
in \cite{7}
\begin{equation}
 W_{\mu\nu} =W_{\mu,\nu} - W_{\nu,\mu}
\end{equation}
Proof of Eq.(12) is given in Appendix-I. An invariant integral $\int W_{\mu\nu} W^{\mu\nu} ~\sqrt{-g}~ d^4x$, and intrinsic geometric property, i. e. Bianchi identity
\begin{equation}
 \partial_\mu W_{\nu\sigma} + \partial_\nu W_{\sigma\mu} + \partial_\sigma W_{\mu\nu} =0
\end{equation}
could also be given in Weyl geometry.

Dirac notes that physicists rejected Weyl unified theory \cite{10}, Weyl himself abandoned it \cite{3, 7}, and Einstein objected to it on 
physical grounds \cite{17}. Dirac was, however fascinated with its simplicity and beauty. Eddington \cite{13} makes a distinction between
natural geometry and world geometry: exact natural geometry applies to real physical world, and world geometry to conceptual or mathematical
space. Eddington is critical of Weyl geometry on mathematical ground \cite{13, 17}. In our view the functions $g_{\mu\nu},~ W_\mu$ characterize
the abstract Weyl space, and it cannot be rejected on either physical or mathematical grounds. The reason is as follows. The Maxwell-like 
equations for the distance curvature $W_{\mu\nu}$ on their own do not represent physics and the gauge-invariant zero length need not be related to light propagation, in that case Eddington's geometry \cite{13} is a natural mathematical generalization of Weyl geometry, and obviously it is not 
a refutation of Weyl geometry. The question now becomes whether Weyl space has a physical realization, that is, does there exist a physical
interpretation of the geometrical quantities $g_{\mu\nu},~ W_\mu$? In analogy to Einstein's interpretation of the coefficiants $g_{\mu\nu}$
of the quadratic groundform (9) representing the gravitational potentials, the coefficients of the linear groundform $W_\mu$ are interpreted
as EM potentials by Weyl. It seems deep influence of Einstein's relativity on Weyl prevented him from propounding the true mathematical
significance of his generalized Riemannian geometry. In fact, Weyl states that a truly realistic mathematics should be conceived in 
line with physics. It may be of interest to reproduce here  intriguing statements due to Weizsaecker in this context \cite{18}: ``A mathematical formalism like, e. g. Hamilton's principle with its mathematical consequences, Maxwell's equations with their solutions, or Hilbert space and 
the Schroedinger equation, is not eo ipso a part of physics. It becomes physics by an interpretation of the mathematical quantities used 
in it, an interpretation which one may call physical semantics.'' It is true that on its own account Maxwell-like equations for $W_{\mu\nu}$
in Weyl geometry do not represent EM phenomena. At the end of the book \cite{18} an interesting discussion, mainly between Linney and 
Weizsaecker, mentions classical and quantum languages, and the problem of continuity, but unfortunately remains confusing and inconclusive.
What is the significance of physical semantics? Does it imply that physics is a language of mathematics?

Translating abstract mathematical reality to physics necessarily involves metaphysical elements, and what one does in practice is to use
the probabilty rules/statistical quantities to get physical quantities. In the Newtonian mechanics the absolute time is a metaphysical
concept without which the role of relative time defined by Newton to explain the observed phenomena would make no physical sense \cite{8}. Mathematically a point, a circle, 2D spherical surface and a punctured 2D Euclidean plane $R^2 -\{0\}$ are well defined. A point is 
dimensionless, a circle with unit diameter has the circumference $\pi$, and 2D spherical surface has the direction holonomy, i. e. the 
direction of a vector parallel transported on the surface after the completion of a cycle is changed. Removing the origin in 2D plane results
into a topological defect in $R^2 -\{0\}$. All the four objects find numerous physics applications, in fact, a point particle is
ubiquitous: classical mechanics, CED, QED and SM. Physical realizations must be approximate, a physical particle is not a point and a hole
in $R^2 -\{0\}$ is not an exact physical void in physical systems like singular vortices in fluid and optics/light beams. According
to Poincare \cite{8} the rule of induction is a mental act exact in mathematics but approximate in physics. Present arguments suggest that
geometry and topology too are approximate in physical world.

The most radical change that we propound is on the nature of time \cite{8, 19}. Absolute time $T$ has a unidirectional flow in discrete steps
and creates the sequence of natural numbers intuitively experienced by mind; the discrete step has an extension that presents itself as
space with each successive step acquiring the property of wholeness as a result of mergence. Since spacetime continuum is an assumption some of the modern quantum gravity theories discard this assumption and  envisage fundamental length and discretized spacetime. Historically Poincare and Weyl did speculate on discrete time \cite{8}, however we postulate discrete nature of time to relate it with the natural numbers. Real numbers are secondary logical and random constructs. The concept of mergence is introduced to define addition in a wholeness: for example, mixing of volumes of water; in a technical sense, superposition of the tails of two Gaussian curves is mergence \cite{8}.   Spatial regions of different numerical magnitudes have random flow in which coalescence and decomposition create connected
quasi-permanent structures that we call matter. The flow has meaning with reference to the absolute time and a common or relative time
variable $t$ can be introduced that measures the changes in the flux of the space. Thus matter has no intrinsic existence but a form of
spatio-temporal bounded structure. In physical geometry a point has a spread and distance between two points has a fluctuating statistical
character. Geometrization of physics becomes a unified picture in which space and time are fundamental.

The ideas of Riemann, Einstein and Weyl on geometrization of physics are entirely different than our ideas. Riemann's infinitesimal geometry is
a great advancement over the finite Euclidean geometry; Weyl terms the idea to gain knowledge from infinitesimal as  the mainspring of
the theory of knowledge, page 92 in \cite{7}. Riemann visualizes metric space only due to the presence of matter, page 98 in \cite{7}.
Einstein's geometrization of macroscopic world is based on the continuation of Riemann's philosophy assuming infinitesimal pseudo-Riemannian
spacetime geometry for gravitation \cite{1, 7}. Weyl follows this line of thinking in the unified theory of gravitation and electromagnetism. 

Clifford's speculation differs from that of Riemann, Einstein and Weyl: matter is a kind of crinkles of space, footnote 2, page 156 in \cite{1}.
Menger's statistical metric space \cite{20} endows statistical distribution function for the distance between any pair of points. Poincare's
observation, see \cite{8}, that $A=B,~ B=C \Rightarrow A=C$ is mathematical while physical experiences show that $A=B,~ B=C \Rightarrow A<C$
has a probability description in the Menger space. Menger suggests that the conceptual problems of microphysics could find resolution in
this underlying geometry. The geometric ideas of Clifford and Menger are of interest in the proposed space and time reality \cite{8, 19, 21},
however they have to be substantially generalized. 

Critique on the relativistic time \cite{8, 19, 21} shows that time in relativity is just a convenient parameter to label/order changes in spatial lengths between two points in 3D space, and the role of vacuum velocity of light is that of a unit conversion factor. Now a physical point in our approach has a statistical spread and fluctuations, therefore, Menger's
statistical metric cannot be used; for the distance between two points also we need a suitable description. Earlier assuming stochastic
electron motion \cite{21} it was suggested that the line element of special relativity represents standard deviation. The proposition that
spatio-temporal objects in the connected whole constitutes physical reality at a fundamental level liberates physics from extraneous matter
\cite{8, 22}. In the stochastic theory \cite{23, 24} one follows the legacy of the Newtonian point particle picture and introduces fluctuations
invoking some kind of hidden thermostat a la de Broglie or Nernst's zero point field (ZPF). The present approach dispenses with them since
spatial regions are themselves in random flux, and spatio-temporal objects have fluctuating spread intrinsic to them. As a consequence point 
particle, instantaneous velocity and potential functions are mathematical idealizations. We seek approximations to define physical properties
from the statistical averages depending on length and time scales.

In special relativity the interpretation of the line element is nicely explained in section 7 of \cite{13}
\begin{equation}
 ds^2 = c^2 dt^2 -dx^2 -dy^2 -dz^2
\end{equation}
\begin{equation}
 (\frac{ds}{dt})^2 =c^2 -v^2
\end{equation}
Time-like interval, i. e. positive value of $ds^2$ is physical for material particles that cannot travel faster than light, i. e. $v<c$
in (16); and $ds^2 =0$ defines null-cone or light geometry. Unfortunately existence of space-like intervals have to be inferred in a contrived 
way if one accepts Eddington's suggestion, moreover physical time is not the common sense time, please see sections 7 and 8 in \cite{13}. Eddington assumes physical matter to have timelike structure and suggests that spacelike particles have different structure. Now spacelike particles could exist only in an instantaneous space, therefore, they are unphysical or impossible structures.  Statistical nature of physical points would suggest that variance or standard deviation could be the best characterization for the metric
\begin{equation}
 ds^2 \rightarrow \sigma^2 =\overline{dl^2} - (\overline{dl})^2
\end{equation}
where $\overline{dl}$ is the average or mean of the distance between two points $\|{\bf r}_2 - {\bf r}_1\|$. A parameter $t$ that represents
time variable with reference to the absolute time $T$ can be introduced using $c$ as a unit conversion factor $\overline{dl^2} = c^2 dt^2$.
Since variance cannot be negative space-like interval has no meaning, and light-cone has zero variance. In the flow of space a set
of physical points having uniform relative change could be selected, and the collection of all such sets $\{I\}$ may be interpreted as
the inertial frames. Assuming one of the sets as a standard for calibration one can define expression (17); time $t$ and light velocity $c$
are convenient parameters. Since changes in spatial relations may occur in both directions the time variable $t$ may have inversion
$t \rightarrow -t$. There is, however no real time reversal since $t$ has meaning with reference to $T$. For more details we refer to \cite{8, 19}.

We have argued that statistical theory has crucial role in translating mathematics to physics; obviously a general and comprehensive 
treatment is needed for this purpose. Here a few examples are worth mentioning. The well known relativistic relation 
\begin{equation}
 E^2 - c^2 {\bf p}.{\bf p} = m^2 c^4
\end{equation}
may be interpreted statistically in which momentum has average value $\overline{\bf p}$ and energy is $c^2\overline{p^2}$. Invariant mass 
turns out to be a measure of standard deviation opposed to its intrinsic invariant value in relativity \cite{13}. Note that in a different way
already in nonrelativistic quantum mechanics Ehrenfest theorem makes use of the expectation values of position, momentum and energy operators.
Infinitesimal Riemannian geometry in physical phenomena has logical interpretation in terms of the Menger metric space \cite{20} while 
pseudo-Riemannian space-time geometry may be obtained based on the generalization of Eq.(17) such that $g_{\mu\nu}$ has statistical significance.
Stationarity of $\int ds$ in relativity corresponds to the minimization of the variance $\sigma$ in Eq.(17). Extending the arguments to CED, the 
EM fields and potentials have statistical nature. One may envisage a true stochastic formulation without hidden thermostat or ZPF \cite{23, 24}
for the Newton-Lorentz equation of motion in this approach.

Preceding discussion leads to a challenging question: Do elementary objects like a single electron and a single photon make physical sense
in the statistical approach? Note that advanced technology and experimental methods have enabled numerous experiments at a single electron and a single
photon level. Electric charge and spin of electron
and spin of photon appear to have unambiguous physical reality. Electron mass in relativity
also has invariant and intrinsic characteristic, please see section 13 in \cite{13}. However, re-examining the relativistic relation written in a different
form
\begin{equation}
 m^2 c^2 = M^2 c^2 -M^2 {\bf v}.{\bf v}
\end{equation}
in view of Eq.(17) statistical interpretation implies that $\sigma^2 =m^2 c^2$ and $\overline{\bf p} =\overline{M \bf v}$. Electron mass $m$
no longer represents intrinsic or internal physical attribute of the electron. Thus charge and spin are the physical quantities that we have to
explain. Could their understanding come from topology? 

Let us briefly explain our idea on the topological defect in 1D. A straight line segment could be continuously deformed to an arc in which shape
and size are unimportant. Removing any one of the interior points of the arc decomposes it inro two parts. A circle cannot be obtained from a straight segment using topological transformation since one needs joining the end points of the segment. Let us consider a real line that is a geometric representation of the set of real numbers
$(R^+, R^-)$: a real number $a$ is positive if $a>0$, and negative if $a<0$. In the real analysis conventionally one divides the line such that 
specifying (not removing) a point O, i. e. the origin, on the line the right is positive and the left is negative; such a line is called a
directed line. There is no continuous transformation that maps points on $R^+ \leftrightarrow R^-$. One may introduce an imaginary axis and 
define a continuous transformation $e^{i\theta}$ on the complex plane to map points on $R^+ \leftrightarrow R^-$. A departure from this construction
\cite{28} envisages the origin as a point defect: real lines on either side have continuous transformations, however crossing the origin
is a discontinuous jump $0^+ - 0^- =\delta_0$, where $\delta_0$ is an infinitesimally small real number. This point defect is in 1D and 
distinct from that of the punctured plane $R^2 - \{ 0 \}$. Tifold in \cite{12} may be related with this kind of topological obstruction.

Topology has two key characteristics \cite{25, 26}, namely continuity and global, and defects and discreteness. The concepts of connectedness,
adjacent points, neighborhoods, and continuous mappings belong to the former while counting of holes/defects to the later; these could be
made technically precise \cite{25}. However metric-independence of topological properties does not preclude the metric spaces from
possessing nontrivial topology. We have argued that ideal mathematical geometry is realized statistically in physics; the statistical metric
spaces may also have nontrivial topology. Holes/defects in physical space-time, for example, vortices and tifold \cite{12, 22} may be such examples.

Could topological models for single electron and single photon be related with the standard field theory? Let us consider the photon model
in which pure gauge potential plays the role for the field description of orbifold and tifold \cite{12}. Now, we face two questions. Firstly
the reality of a discrete localized photon has been questioned in the literature, e. g. Lamb's anti-photon viewpoint \cite{27} and
in stochastic electrodynamics \cite{24}. However, photodetectors do show direct evidence of an observed photon \cite{28}. Moreover, in
the SM along with weak gauge bosons photon is also a gauge boson and the observed data in particle physics experiments \cite{29} show
the presence of photon directly as compared to rather indirect observation of weak gauge bosons inferred from their decay modes. Low
frequence EM radiation and electron scattering treated classically explains the measured cross sections satisfactorily using the 
Thomson formula. An important concept of classical electron charge radius $e^2/m c^2$ also follows from this. On the other hand, at high energy
$\nu \approx m c^2 /h$ quantum theory of Compton scattering utilizing photon picture becomes essential. For more details we refer to
\cite{16} and review on the photon concept in \cite{30}.

The second question is an age-long one: potentials are convenient calculational tools or they have independent reality. Aharonov and Bohm 
\cite{4} state the problem very clearly in the modern context that potentials seem to have the role of auxiliary variables devoid of physical
significance either because they can be eliminated from the equations in classical theory (CED) or due to the principle of gauge invariance in quantum
theory. Vast literature on this contentious issue seems to ignore the important fact that the observed photon is a spin one gauge boson in SM
\cite{29} and spacetime symmetries associate spin one with a vector field not a tensor field. Does it not prove the reality of EM potentials?

Originally the Maxwell equations were obtained from the macroscopic EM phenomena. There is no compelling reason either experimental or
theoretical to associate EM fields with a single photon; it is just a convention. Traditional meaning of the principle of gauge invariance
has to be modified. A logically consistent approach could be to distinguish a single photon and a system of large number of photons 
described by statistically averaged physical quantities. Postulating ${\bf E} = {\bf B} = 0$ for a single photon has profound implication
on the physical nature of photon and the significance of EM potentials \cite{12}. Denoting the EM potentials for a single photon by 
$\mathcal{A}_\mu$ we have the following equations satisfied by them
\begin{equation}
 {\bf \nabla}.\mathcal{ A} +\frac{1}{c} \frac{\partial \mathcal{A}_0}{\partial t} = 0
\end{equation}
\begin{equation}
 {\bf \nabla} \mathcal{A}_0 + \frac{1}{c} \frac{\partial \mathcal{A}}{ \partial t} =0
\end{equation}
\begin{equation}
 {\bf \nabla} \times \mathcal{A} =0
\end{equation}
First-order partial differential Eqs. (20)-(22) may be combined to derive the second-order wave equations
\begin{equation}
 \nabla^2 \mathcal{A} -\frac{1}{c^2} \frac{\partial^2 \mathcal{A}}{\partial t^2} =0
\end{equation}
\begin{equation}
 \nabla^2 \mathcal{A}_0 -\frac{1}{c^2} \frac{\partial^2 \mathcal{A}_0}{\partial t^2} =0
\end{equation}

Lorentz gauge (20) has the nice property that it is manifestly Lorentz covariant, moreover it has formal resemblence with the continuity equation for fluid flow. For a single photon 4-vector potential itself is interpreted as energy-momentum 4-vector. The electric charge unit $e$ has implicit presence in the definition of EM potentials expressed in terms of charge and current density. Now factoring out electric charge unit $e$ from the EM fields and potentials one gets the geometrical unit for them; in Weyl geometry a geometric unit arises naturally. Multiplication with $\hbar$ makes the vector potential $\mathcal{A}$ to have the dimension of momentum.
We have proposed a screw disclination and tifold model for photon \cite{12} in which spin is a topological invariant. Particle nature of
photon has been suggested to be due to the topological defect associated with the spin.

In an abstract mathematical formalism Salingaros \cite{31} obtains holomorphic field equations in 4D spacetime for two types of tensors
that formally resemble Maxwell equations. For a vector field $a_\mu$ the holomorphy condition is the set of equations (18) in \cite{31}.
These equations are nothing but Eqs. (20)-(22) as above. The author emphasizes the spacetime interpretation of EM phenomena; this
interpretation is in the spirit of Weyl's interpretation in which EM field tensor is a spacetime distance curvature. Unfortunately,
the equations satisfied by $a_\mu$ have been unimaginatively treated as just the Lorentz gauge condition and the identity that
${\bf E} ={\bf B} =0$. Note that the equations for $a_\mu$ in \cite{31} are independent of Maxwell-like equations for a tensor of type 2,
and could be viewed as generalized Cauchy-Riemann equations. For this reason, mathematically the equations for $a_\mu$ are as significant
in 4D as Cauchy-Riemann equations are in 2D. The photon model based on Eqs. (20)-(22) in \cite{12} acquires added significance.

The present perspective on mathematics and physics, specifically in the context of electromagnetism, leads to two new directions: 1) foundations
of CED have to be re-examined interpreting EM fields as statistically averaged macroscopic quantities for photon fluid, and 2) the 
fundamental role of pure gauge potentials in the topology necessitates a thorough analysis of the gauge conditions in Weyl \cite{7, 13} and Weyl-Dirac \cite{10} theories.

\section{\bf Gauge invariance and Weyl geometry: new insights}

Relativity and 4D spacetime geometry had profound influence on Weyl, and his aim of a unified theory was greatly inspired by Mie's theory
\cite{7}. Curiously Weyl perceptively quotes Clifford's speculation of 1875:``the theory of space curvature hints at a possibility of
describing matter and motion in terms of extension only''. Weyl asserts that ``physics does not extend beyond geometry'', but reverts to
the conventional setup of matter immersed in space in his unified theory. The central problem in his work is that of the electron model; according to him the Maxwell-Lorentz theory is invalid in the interior of electron. Unfortunately the unified theory of Weyl could not make much progress in this objective \cite{7}. Relativity of magnitude together with the relativity of motion resulted into a true infinitesimal geometry, namely
Weyl geometry: a point in spacetime needs specification of coordinate system and a gauge or units of measure.  Besides \cite{7, 13} a self-contained necessary tensor analysis in Weyl geometry could be found in
Dirac's paper \cite{10}.

Following the metric and gauge conventions in \cite{10, 13} in-invariants and in-tensors are gauge invariant. Metric tensor $g_{\mu\nu}$ is
a co-tensor of power 2, and $\sqrt{-g}$ has power 4. The generalized affine connection or Christoffel symbol invariant under gauge
transformation denoted by $^*\Gamma^\alpha_{~\mu\nu}$ is 
\begin{equation}
 ^*\Gamma^\alpha_{~\mu\nu} =\Gamma^\alpha_{~\mu\nu} - g^\alpha_\mu A_\nu - g^\alpha_\nu A_\mu +g_{\mu\nu} A^\alpha
\end{equation}
In Weyl geometry the transformation of the tensor quantities has to be considered for both coordinate and gauge transformations, therefore,
one has to assign Weyl powers as well to them. The meaning of tensor densities and the role of $\sqrt{-g}$ could be found in sections (16) and (17)
in \cite{7}. Following Eddington \cite{13} the scalar curvature $^*R$ is termed a co-invariant of power $-2$
\begin{equation}
^*R =R -6 W^\mu_{~:\mu} + 6 W^\mu W_\mu
\end{equation}
Assuming that variational principle for an appropriate action integral yields the field equations for the unified theory the problem becomes
that of constructing $W$ in the action
\begin{equation}
 S =\int W \sqrt{-g} ~d^4x
\end{equation}
Gauge invariance of the action $S$ demands that $W$ must be a co-scalar of power $-4$ since $\sqrt{-g}$ has power 4. Weyl opts for the 
squared scalar curvature $^*R^2$, however he admits that it may not be realized in nature. Dirac \cite{10} introduces an arbitrary scalar
function $\beta$ of power $-1$  and constructs $W$ linear in $^*R$. Their respective functions are 
\begin{equation}
 W_W = ~^*R^2 - \alpha ~ W^{\mu\nu} W_{\mu\nu}
\end{equation}
\begin{equation}
 W_D = \beta^2~ ^*R - \frac{1}{4} F^{\mu\nu} F_{\mu\nu}
\end{equation}
Distinct symbols in (28) and (29) for the second rank antisymmetric tensor signify the fact that in Weyl action it is a geometric quantity, namely the distance curvature, while Dirac proceeds with the assumption that it represents the EM field tensor.

Here our main interest is regarding the issue of gauge invariance rather than the field equations. In Weyl geometry the fundamental postulate 
is that the comparison of lengths at a distance is not possible and a unique and uniform calibration does not exist. To reconcile this with the 
physical experiences Weyl puts forward two arguments that have been critically elaborated by Eddington. The first argument is that we may
define a unit of length assuming
\begin{equation}
 ^*R = 4\lambda
\end{equation}
where $\lambda$ is a constant. This gauging equation is supposed to represent a natural gauge. The existence of the natural gauge could be 
justified in analogy to the existence of a unique Galilean system of coordinates in the pseudo-Riemannian spacetime geometry of general
relativity. However, there arises a serious difficulty in fixing a gauge since the law of parallel displacement of a vector contradicts it.
To resolve this problem the second argument is developed based on the philosophy that there are two ways of determining the quantities:
by persistence and by adjustment. A priori one cannot conclude that pure transference following the tendency of persistence is integrable.
Size of an electron is determined by adjustment in view of the presence of space curvature, and not by the persistence of its time history.
Another example is that of electric charge: conservation of charge cannot explain why electron has the same charge after a lapse of long 
time, and why this charge is the same for all electrons. Thus charge is determined by adjustment, not by persistence, i. e. at
every instant of time the state of equilibrium of negative electricity adjusts to this value. In Eddington's view this interpretation of
the gauge principle amounts to a graphical representation of physical facts; a measuring rod taken from one point to another for 
calibration could be taken to satisfy the natural gauge condition (30), and this process should not be viewed as a parallel displacement of a vector.

The question is: What is the significance of the gauge condition (30) in the unified field equations? First an important point has to be noted
that even if $^*R$ is not constant one could change the measure to transform it to a constant in a new gauge. Though unified field equations
correspond to Einstein-Maxwell like field equations; let us drop ``like'' and accept Weyl's interpretation that they represent electromagnetism
and gravitation. In that case the Maxwell equations for $W_{\mu\nu}$ depart radically from standard CED equations since the gauge potential
$W_\mu$ itself acquires the role of source current density
\begin{equation}
 J_\mu \propto W_\mu
\end{equation}
Weyl postulates a unique calibration setting $^*R=1$ and introduces the radius of curvature of the universe for the recalibration. This leads to the gauge condition (30)  and $\lambda$ acquires the significance as a cosmological constant in the Einstein equation.

A remarkably unusual result (31) that charge current density is equivalent to the potentials has intriguing aspects. First it may be noted that the current continuity equation follows from the Maxwell equation
\begin{equation}
 J^\mu_{~:\mu} =0
\end{equation}
In view of Eq.(31) formally we have
\begin{equation}
 W^\mu _{~:\mu} =0
\end{equation}
Lorentz gauge condition of the standard CED assumes a radically new role in Eq.(33): it is inseparably related with the charge conservation law and according to Weyl, Eq.(31) shows that the electric charge is ``diffused throughout the world'' since the gauge potential is present in all space.

Alternative derivation of Eq.(32) is based on the contraction of the Einstein equation and the use of the gauge condition (30). Two routes to get current continuity pose a serious problem since the Lorentz gauge condition is inconsistent with the gauge condition (30). Weyl \cite{7} and Eddington \cite{13} seek a resolution of this issue exploring the interplay between the energy tensor of matter and that of EM field. The arguments run as follows. Total energy-momentum tensor in the Einstein field equation with a cosmological constant could be assumed comprising of the EM part, $E_{\mu\nu}$, and the matter one $M_{\mu\nu}$. The contraction gives the trace equation interpreted in terms of the proper-density of matter
\begin{equation}
 \rho_M \propto ~ R- 4\lambda
\end{equation}
Using the gauge condition (30) $\rho_M$ becomes
\begin{equation}
 \rho_M \propto ~ W^\mu_{~:\mu} -W^\mu W_\mu
\end{equation}
Empty space is defined to be the absence of electrons, though EM field is nonzero, then $\rho_M = 0$, and Eq.(35) implies
\begin{equation}
 W^\mu W_\mu = W^\mu_{~:\mu}
\end{equation}
It is clear that the problem has become more intricate as there are now three incompatible gauge conditions: natural gauge (30), Lorentz gauge (33), and nonlinear gauge (36). Weyl asserts that the relation (31) has fundamental importance, in that case, there is no empty space and the Lorentz gauge would have general validity. However substituting (33) in Eq.(35) one has
\begin{equation}
 \rho_M \propto W^\mu W_\mu ~\Rightarrow ~ \rho_M \propto ~ -J^\mu J_\mu
\end{equation}
The positivity of mass density implies that $J_\mu$ is a spacelike vector, and it cannot be true for the electric charge. A speculative idea is to consider interior of the electron where Eq.(37) holds, and the outside region separated by the boundary of the electron. Eddington suggests that the internal structure of the electron may have the constituents something like magnetic charges. In Weyl theory the Lorentz gauge condition occupies special position, and is assumed to hold in general. On the other hand, the nonlinear gauge (36) is interpreted to represent the vanishing of the current density that in view of (31) becomes consistent with the Lorentz gauge. Outer region of the electron, however has small charge and current extending to infinity.

Preceding account shows that the issue of gauge invariance in the context of the Weyl action function for the unified theory could not be settled satisfactorily. Of course, Weyl himself states that, ``I do not insist that it is realized in nature'' \cite{7}, and Eddington \cite{13} remarks that Weyl action ``has no deep significance''. We emphasize that questioning the specific action function, e. g. Eq.(28), does not mean a rejection of Weyl geometry. For this reason, Dirac's view on Weyl geometry is very important: wider group of spacetime transformations for physical laws noted in the conclusion of his paper \cite{10}. It is important to note that Eddington's criticism is applicable to Dirac's action function (29): combining two in-invariants of different forms, namely the Maxwell action 
$F^{\mu\nu} F_{\mu\nu}$ and $\beta^2 ~^*R$.

Dirac revived Weyl geometry \cite{10} primarily with the motivation of his large number hyp[othesis in which gravitational constant varies with time. He assumes a constraint equation
\begin{equation}
 ^*R =0
\end{equation}
and using $\beta^2$ as a Lagrange multiplier constructs a gauge invariant action integral (29). Dirac lists three choices of the gauge: 1) the natural gauge; if $W_{\mu\nu} =0$ one has $W_\mu =0$, 2) the Einstein gauge $\beta=1$ in which Einstein field equations are obtained for the vanishing of the EM quantities, and 3) the atomic gauge. Surprisingly Dirac makes no attempt to explain (38) in relation to Weyl's natural gauge (30). The issue of gauge invariance in empty space with the contradicting gauge conditions (33) and (36) that troubled Weyl and Eddington in connection with CED and electron structure also did not receive Dirac's attention. Though symmetry breakdown in Weyl space for charge conjugation and time reversal is separately pointed out by him, Weyl's speculation that inequality of positive and negative electricity may be related with the past $\rightarrow$ future asymmetry, see pp 310-311 in \cite{7}, is also not discussed.

Weyl in the Preface to the American edition of \cite{7} states that his attempt of a unified theory had failed. It is well known that Weyl's original idea of gauge symmetry metamorphosed into the modern gauge field theories \cite{2, 3}. Revival of the original Weyl geometry by Dirac \cite{10} and current interest in this geometry \cite{9} are mainly in connection with the problems in cosmology. The literature on Weyl theory has not unraveled the deeper issues related with gauge invariance and gauge conditions. Could there be a new strand for Weyl geometry applicable to particle physics? This question is made precise in the light of the new perspective emerging from our analysis.

{\bf [I]} Weyl and later Eddington exclusively focused on electron and diffused charge interpretation. They did not explore possible implications on the nature of photon and EM fields. Note that Einstein's light quantum hypothesis was well known at that time though the word photon was coined later in 1926. It is also not clear as to why Dirac in spite of his sustained engagement with new theories of the electron did not investigate Weyl geometry from this angle. May be it was due to his preoccupation with the large number hypothesis in \cite{10}. Setting an arbitrary constant equal to 6 appearing in his action integral he obtained vacuum field equations. Unfortunately that missed uncovering geometric origin of current density in the Maxwell equation. This issue is of fundamental importance for physical interpretation of the EM potentials and current density. To see this let us compare the standard Maxwell-Lorentz theory and Weyl theory. Eq.(7) shows that vanishing of EM field tensor does not necessarily imply vanishing of the EM potentials since one may have $A_\mu = \partial_\mu \chi$ and $F_{\mu\nu}=0$. In contrast, in Weyl theory the current density (31) severely restricts the potentials since $W_{\mu\nu} =0 ~ \Rightarrow ~ W_\mu =0$. As a consequence the freedom to envisage multi-valued pure gauge potential in Weyl theory is lost.

{\bf [II]} Relaxing Dirac's assumption of setting the constant equal to 6 Papini and his collaborators have extensively studied Weyl-Dirac theory \cite{32}. However their work does not throw new light on the gauging conditions, and merely repeat Dirac's argument that to reconcile Weyl gauge principle with the atomic standrads of length one could postulate two metrics. Papini's idea on multi-valued scalar field is however quite interesting.

{\bf [III] } Weyl's motivation for his geometry was unified theory for gravitation and electromagnetism \cite{7}. The subject as such has three aspects: mathematical comprising of geometry, physical interpretation of geometric quantities and field equations, and physics motivated action principle. Most of the criticism pertain to Weyl action (28) and Dirac too modifies the action to (29) in \cite{10} introducing an additional field variable as a Lagrange multiplier in the assumed constraint equation (38). For physical interpretation, analogy is drawn to Einstein's theory of gravitation. Einstein field equation is founded on the identification of the metric tensor as gravitational potential and the equivalence between the Einstein tensor and energy-momentum tensor. Now, Einstein field equation could be derived from the Hilbert-Einstein action. But, in general, a symmetry of the action may not necessarily be a symmetry of the field equation. For a lucid exposition on the calibration/gauge symmetry we refer to Bock \cite{33}.

Axiomatically the metric tensor $g_{\mu\nu}$ for the quadratic form (line element) and $W_\mu$ for the linear form are the fundamental geometric quantities in Weyl space. It is well known, see e. g. \cite{13} that in Riemannian geometry one can construct Galilean coordinate system for constant $g_{\mu\nu}$, and the necessary and sufficient condition for flat geometry is that Riemann curvature tensor is zero. Note that the Christoffel symbol vanishes for constant metric tensor, however it is not a tensor, therefore, it may assume a nonvanishing value in a new coordinate system. In Weyl geometry the definition of the distance curvature $W_{\mu\nu}$ ensures that it vanishes if $W_\mu =0$, however, for $W_{\mu\nu} =0$ one could have a nonvanishing pure gradient for $W_\mu$. It may be argued that by a gauge transformation $W_\mu$ could be made equal to zero. In a nontrivial topology of Weyl space an interesting possibility exists: a nonzero $W_{\mu\nu}$ exists in a small localized region and it is zero everywhere outside that region, but $W_\mu$ is nonzero there. This case is similar to the topology due to a confined magnetic flux in a solenoid for AB effect \cite{4}.

Concluding this section, a new physical manifestation of Weyl space is indicated from our analysis where one may consider two ideas: a single electron and a single photon could be visualized as topological defects in Weyl space, and EM fields and current density are treated as macroscopic quantities having statistically averaged significance. In the light of the proposition of statistical metric tensor discussed in section II it seems natural to associate statistical meaning to the gauge transformation.

\section{\bf New Field Equations in Weyl Space: Nontrivial Topology}

Rather than Dirac's emphasis on the large scale structure of the universe in connection with the Weyl geometry, a new action function was setup
to explore the nature of electron charge in \cite{11}. Let us call it a generalized Weyl-Dirac theory. A beautiful feature of Weyl geometry is the gauge invariance of zero length:
massless fields become natural objects in this geometry. If one speculates that at a fundamental level electron comprises of massless fields
then photon and electron assume special place in Weyl geometry. Could there be a fundamental role of nontrivial topology and the gauge
potentials in this scenario? To address this question we discuss the generalized Weyl-Dirac theory.

Introducing an in-invariant field $\Psi$ representing electron in the proposed action integral \cite{11} we have
\begin{equation}
 S = \int (~^*R \xi + p F^{\mu\nu} ~^*R_{\mu\nu} ) \sqrt{-g} ~ d^4x
\end{equation}
\begin{equation}
 \xi = \Psi^{:\mu} \Psi_{,\mu}
\end{equation}
Field $\xi$ is a co-scalar of power -2 and it seems akin to Dirac's $\beta^2$, however unlike cosmological interpretation of $\beta$, here
the field $\xi$ is proposed to be related with the electron and photon. Variational principle treating $g_{\mu\nu}, A_\mu, \Psi$ as independent dynamical variables
leads to a set of field equations comprising of the gravitational field equation, modified Maxwell field equation, and a field equation for 
$\Psi$, i. e. Equations (25)-(27) respectively in \cite{11}. The expression for $^*R_{\mu\nu}$ contains antisymmetric tensor $F_{\mu\nu}$
following \cite{13} in contrast to the symmetrized $^*R_{\mu\nu}$ used by Dirac \cite{10}. We have returned to the standard notation 
$A_\mu,  F_{\mu\nu}$ for the EM quantities but expressed in geometrical units. The source current density in the modified Maxwell equation,
setting the arbitrary constant $p=1/4$, shows dependence on the gauge potentials
\begin{equation}
 F^{\mu\nu}_{~;\nu} = - 3(\xi^{:\mu} +2\xi A^\mu)
\end{equation}
The derivation of this equation is presented in Appendix-II.
The conservation law for the current density
\begin{equation}
 J^\mu =- 3(\xi^{:\mu} +2\xi A^\mu)
\end{equation}
follows from Eq.(41); it could also be obtained as an identity from the contraction of the Einstein field equation (25) in \cite{11}. Thus we have
\begin{equation}
 \xi^{:\mu}_{~:\mu} + 2 \xi_{,\mu} A^\mu + 2 \xi A^\mu_{~:\mu} = 0
\end{equation}

The modified Maxwell field equation (41)  differs from that of Weyl \cite{7} and admits nonvanishing EM potantial for $F_{\mu\nu} =0$ given by
\begin{equation}
 \xi^{:\mu} + 2 \xi A^\mu =0
\end{equation}
Equation (44) needs a careful analysis since $\xi$ is also related with the postulated electron field $\Psi$ in Eq.(40). Let us consider the full set of the field equations and assume that the EM fields are zero then the energy tensor $E_{\mu\nu}=0$. Making use of Eq.(44) to eliminate derivatives of $\xi$ the Einstein field equation becomes
\begin{equation}
 G^{\mu\nu} = - \frac{^*R}{\xi} \Psi^{;\mu} \Psi^{;\nu} +A^\alpha A_\alpha g^{\mu\nu} +2 A^\mu A^\nu +2 A^{\mu ;\nu} - 2 A^\alpha_{~:\alpha} g^{\mu\nu}
\end{equation}
where the Einstein tensor is
\begin{equation}
 G^{\mu\nu} = R^{\mu\nu} -\frac{1}{2} g^{\mu\nu} R
\end{equation}
Note that the trace of the Einstein field equation (45) gives just the identity (26). Imposing Weyl natural gauge condition (30) the $\Psi$ field satisfies the massless wave equation
\begin{equation}
 \Psi^{;\mu}_{~:\mu} =0
\end{equation}
Introducing the energy tensor for $\Psi$ and $A_\mu$
\begin{equation}
 T^{\mu\nu}_{(\Psi)}  =\Psi^{;\mu} Psi^{;\nu} -\frac{1}{2} g^{\mu\nu} \Psi^{;\alpha}\Psi_{,\alpha}
\end{equation}
\begin{equation}
 T^{\mu\nu}_{(A_\alpha)} = -A^\alpha A_\alpha g^{\mu\nu} - 2 A^\mu A^\nu -2 A^{\mu;\nu} +2 A^\alpha_{~:\alpha} g^{\mu\nu}
\end{equation}
Eq.(45) is re-written as
\begin{equation}
G^{\mu\nu} + 2 \lambda g^{\mu\nu} = -\frac{4 \lambda}{\xi} T^{\mu\nu}_{(\Psi)} -T^{\mu\nu}_{(A_\alpha)} 
\end{equation}
A simple calculation gives the trace expressions
\begin{equation}
 T_{(\Psi)} = -\xi
\end{equation}
\begin{equation}
 T_{(A_\alpha)} = -6 A^\alpha A_\alpha + 6 A^\alpha_{~:\alpha}
\end{equation}
One of the important consequences of the set of the field equations is that both the natural gauge condition (30) and the Lorentz gauge condition (33) or (20) could be simultaneously assumed without any inconsistency. Making this assumption we have
\begin{equation}
 T^{\mu\nu}_{(A_\alpha)} = -A^\alpha A_\alpha g^{\mu\nu} - 2 A^\mu A^\nu -2 A^{\mu;\nu}
\end{equation}
\begin{equation}
 T_{(A_\alpha)} = -6 A^\alpha A_\alpha
\end{equation}
\begin{equation}
 T_{(A_\alpha)} = R- 4\lambda
\end{equation}

The expression for the energy tensor (53) contains only the vector gauge field and logically one would expect its interpretation as that of a massless vector field. In relativistic field theory it is known that for a massive vector field the representation splits into a 3-vector and a scalar under rotation, and the Lorentz gauge condition is needed to resolve the question of the positivity of energy \cite{34}. In our case, the interpretation of (53) as energy tensor of $A_\alpha$ is inferred from the Einstein field equation (50). It is important to note that the trace expressions (52) and (54) remain unaltered even if EM fields are nonzero since $E_{\mu\nu}$ is traceless. On the other hand, the energy expression (90.6) in Weyl theory \cite{13} containing potentials is attributed to the matter energy tensor. Recall that in the standard CED Eqs. (20)-(24) show that for EM field free case, and the assumption of the Lorentz gauge, each component of $A_\mu$ satisfies the massless wave equation. Field theoretic arguments \cite{34} and topological  model of photon \cite{12} suggest that pure gauge field $A_\mu$ may be identified as photon field.

The presence of $\xi$ in Eq.(44) and Eq.(50) indicates subtle role of the coupling between electron field $\Psi$ and photon field $A_\mu$. Is it possible to decouple electron from the influence of the EM field as well as pure gauge potential? Towards this aim let us examine the consequences of setting $\xi = 0$ . Eq.(40) shows that 
\begin{equation}
 \Psi ^{;\mu}\Psi_{,\mu} =0
\end{equation}
Eq.(50) reduces to 
\begin{equation}
 4 \lambda \Psi^{;\mu} \Psi^{;\nu} =0
\end{equation}
Mathematically there is no contradiction if the expression multiplied by by $\xi$ in Eq.(50) is also equal to zero. In that case, one gets the Einstein equation 
\begin{equation}
 G^{\mu\nu} = - T^{\mu\nu}_{(A_\alpha)}
\end{equation}
Physical arguments show that one should also take $\lambda =0$ as a solution of Eq.(57) and 
\begin{equation}
 \Psi^{;\mu} \Psi^{;\nu} \neq 0
\end{equation}
Trace of Eq.(58) shows that $^*R =0$ in agreement with $\lambda =0$ . Eq.(55) reduces to 
\begin{equation}
 R = - 6 A^\alpha A_\alpha
\end{equation}

Decoupled electron field $\Psi$ satisfies the massless wave equation (47) as well as Eq.(56). Multiplying Eq.(47) by $\Psi$ and integrating over a compact manifold without boundary gives an uninteresting result when use is made of Eq.(56). Evidently $\Psi$ field cannot be a continuous field of spacetime variables. Let us examine the possibility of a discontinuous solution. In the following for most of the discussions on topological aspects we consider the illustrative examples in flat spacetime geometry, and conjecture that the topological characteristics would be carried over to the Weyl space. The conjecture is motivated by metric-independence of topological properties.

Define a boundary surface at which the field is discontinuous
\begin{equation}
 F(x_\mu) = L(x, y, z) -ct =0
\end{equation}
Instead of a plane wave solution of Eq.(47) the field on the discontinuity surface could be assumed as
\begin{equation}
 \Psi = e^{i L - i \omega t}
\end{equation}
Substituting expression (62) in the wave equation, and making the approximation that $\nabla^2 L$ is small compared to other terms we get
\begin{equation}
 {\bf \nabla} L.{\bf \nabla} L =\frac{\omega^2}{c^2}
\end{equation}
Using (62) in Eq.(56) we get once again Eq.(63). Following a nice discussion in the Appendix VI of \cite{35} we define a unit normal vector to the surface (61)
\begin{equation}
 \hat{\bf n} = \frac{{\bf \nabla} F}{|{\bf \nabla}F|}
\end{equation}
and the speed of the moving discontinuity by
\begin{equation}
 v_f = -\frac{1}{|{\bf \nabla}F|} \frac{\partial F}{\partial t}
\end{equation}
In the present case the discontinuity moves at the velocity of light. In analogy to geometrical optics limit of the light wave propagation \cite{35}, Eq.(56) or Eq.(63) is the eikonal equation. Thus, the electron field is a defect propagating at the speed of light.

In the Einstein field equation (45) the energy contribution of $\Psi$ field disappears, and the decoupled electron has no self-field of the EM origin. The significance of this field is two-fold: its existence signals a nontrivial topology of Weyl geometry, and the possibility of nonzero $\xi$ makes it observable through EM interaction via Eq.(42). The drawback of scalar field representation is that the spin property is not explained, in fact, it corresponds to spinless electron in CED; for a recent attempt to explore spin of electron, see \cite{22}.

Decoupled photon field satisfies Eq.(58) and Eq.(60). Assuming traceless energy-momentum tensor for $A_\mu$ Eq.(58) leads to $R=0$. In this case Eq.(60) reduces to the nonlinear gauge 
\begin{equation}
 A_\alpha A^\alpha =0
\end{equation}
Mathematically Eq.(66) shows that $A_\alpha$ is self-perpendicular. The most important result is that the gauge conditions are unambiguously consistent and compatible: $\lambda =0 ~ \Rightarrow ~ ^*R =0$; each term in $^*R$ vanishes, therefore $R=0$, Lorentz gauge (33) and nonlinear gauge (66) are also satisfied. Note that for a pure gauge field , $F_{\mu\nu} =0$ and using the Lorentz condition one finds that $A_\mu$ satisfies the massless vector wave equation. To unravel the topology of Weyl geometry we examine the nature of $A_\mu$ in three ways.

{\bf I ~} Simplest and conventional solution is given by 
\begin{equation}
 A_\mu = \eta_{,\mu}
\end{equation}
Substituting (67) in the Lorentz condition we have 
\begin{equation}
 \eta^{;\mu}_{~:\mu} =0
\end{equation}
and the nonlinear gauge (66) becomes
\begin{equation}
 \eta^{;\mu} \eta_{,\mu} =0
\end{equation}
Eqs.(68) and (69) for $\eta$ are exactly the same as the ones for $\Psi$, therefore, for a nonvanishing $\eta$ the propagating discontinuity interpretation would be essential. The photon field becomes a propagating topological defect.

It is, however not necessary to proceed with the solution (67); the field $A_\mu$ itself shows the topological property if we consider the discontinuous surface (61) and calculate $A_\mu$ on it following \cite{35}. The vector potential ${\bf A}$ and the scalar potential $\Phi$ may be defined on either sides of the surface (61) and using the unit step or Heaviside function we may write, for example, for the vector potential 
\begin{equation}
 {\bf A} = H(-F) {\bf A}^1 + H(F) {\bf A}^2
\end{equation}
The derivative of the step function is the Dirac delta function, therefore space and time derivatives of the potentials can be calculated using the standard calculus. For example,
\begin{equation}
 {\bf \nabla} \times {\bf A} = H(-F) {\bf \nabla} \times {\bf A}^1 +H(F) {\bf \nabla} \times {\bf A}^2 +\delta(F) {\bf \nabla} F \times \Delta {\bf A}
\end{equation}
\begin{equation}
 \frac{\partial {\bf A}}{\partial t} =H(-F) \frac{\partial {\bf A}^1}{\partial t} +H(F) \frac{\partial {\bf A}^2}{\partial t} + \delta(F) \frac{\partial F}{\partial t} \Delta {\bf A}
\end{equation}
\begin{equation}
 \Delta {\bf A} = {\bf A}^2 - {\bf A}^1
\end{equation}
Expression(73) denotes the discontinuous change. The Lorentz condition using (64) and (65) finally leads to 
\begin{equation}
 \hat{\bf n} .\Delta {\bf A} - \Delta \Phi =0
\end{equation}
Note that ${\bf E} =0$ and ${\bf B}=0$ also give Eq.(74). On the moving surface the potentials are calculated using the fact that ${\bf A}^2,\Phi^2$ are zero, therefore, Eq.(74) becomes
\begin{equation}
 \hat{\bf n} .{\bf A} - \Phi =0
\end{equation}
It is easy to verify that in view of Eq.(75) the nonlinear gauge (66) is satisfied. Thus the photon is a discontinuity of the potentials propagating at the speed of light.

{\bf II ~} The field theoretic description of topological photon \cite{12} is given based on $F_{\mu\nu}=0$ and the Lorentz condition. Here in addition we have the constraint (66). Screw disclination for the vector field ${\bf A}$ has been interpreted in terms of 2D defect termed orbifold and 1D defect as tifold in \cite{12}. The 4-vector field $A_\mu$ is split into transverse ${\bf A}_t$ and longitudinal components $(A_z, \Phi)$ choosing a propagation direction along z-axis. Following \cite{36} the assumed solution in complex representation is 
\begin{equation}
 A_x = k r e^{i\chi}
\end{equation}
\begin{equation}
 A_y = i k r e^{i\chi}
\end{equation}
where $\chi = \theta +k z -\omega t$. It is easy to verify that 
\begin{equation}
 \frac{\partial A_x}{\partial x} + \frac{\partial A_y}{\partial y}=0
\end{equation}
and the Lorentz condition reduces to 
\begin{equation}
 \frac{\partial A_z}{\partial z} +\frac{1}{c} \frac{\partial \Phi}{\partial t} =0
\end{equation}
Could one find $A_z, \Phi$ such that for the solutions (76) and (77) the nonlinear gauge (66) is also satisfied? Since the sum of the squares of the real parts of (76) and (77) is $k^2 r^2$ a simple solution $A_z = kz$ and $\Phi = -\omega t$ leads to a curious result: Eq.(66) becomes the null-cone equation
\begin{equation}
 x^2 + y^2 + z^2 -c^2 t^2 =0
\end{equation}
Note that Nye \cite{36} considers the electric field vector to study disclination for polarization effects of EM wave; in that case ${\bf \nabla}.{\bf E} =0$. Here we consider $A_\mu$ that satisfies the Lorentz condition, and hence follows Eq.(79).

{\bf III ~} Definition of $F_{\mu\nu}$ , expression (5) . in analogy to curl in 3D could be treated as curl of $A_\mu$ in 4D , see section 3.2 in \cite{13}. The 4-divergence of $A_\mu$ is $\partial^\mu A_\mu$. Now the curl-free and divergence-free vector in 3D may have peculiar characteristics made transparent in the abstract language of differential forms and de Rham periods. An intuitive and physics-oriented account on de Rham periods on the manifolds can be found in \cite{37}; also summarized in \cite{12}. In a Euclidean domain formally a 1-form is decomposed into three parts : an exact form (nonzero divergence, zero curl) , a closed form (zero divergence, nonzero curl), and a harmonic form (both divergence and curl are zero). According to de Rham theorem the closed loop integral of the harmonic form over a non-bounding cycle counts the number of holes/topological defects. It may be asked if de Rham theorem could be extended to 4D spacetime.

Formally photon field, i. e. $A_\mu$ is curl-free $F_{\mu\nu}=0$ and divergenceless $\partial^\mu A_\mu =0$; it implies that $A_\mu dx^\mu$ is a harmonic form with the loop integral
\begin{equation}
 \oint A_\mu dx^\mu = 2\pi N
\end{equation}
Note that in Weyl geometry $A_\mu$ has the dimension of $length^{-1}$. Here N is an integer counting the number of topological defects, and for a single photon $N=1$. Weyl geometry for harmonic 1-form (81) acquires a nontrivial topology. However, the non-Euclidean spacetime geometry limits the straightforward application of de Rham theorem. We may get some physical insight considering the vector potential ${\bf A}$ such that ${\bf B} =0$, and take the manifold $R^3 -\{ 0\}$. An example of a harmonic form is given in Appendix C of \cite{37}
\begin{equation}
 {\bf A} = \frac{y}{r^2} \hat{\bf i} -\frac{x}{r^2} \hat{\bf j}
\end{equation}
Though ${\bf \nabla}.{\bf A} =0$ and ${\bf \nabla} \times {\bf A}=0$, the loop integral enclosing the origin is nonzero
\begin{equation}
 \oint {\bf A}.{\bf dl} =2\pi
\end{equation}
The harmonic form (82) may also be given a physical realization in the form of a singular vortex: the fluid flow in concentric circles around the origin at which one has a singularity of the velocity field. Note that multiplying $A_\mu$ by $\hbar$ its dimension is that of momentum, and Eq.(83) may be related with the spin of the photon \cite{30}.

In this section possible nontrivial topology of the Weyl space has been investigated based on the generalized Weyl-Dirac theory \cite{11}. The most important new result is that $\Psi$ field for spinless electron and $A_\mu$ for photon represent topological defects in space-time of Weyl geometry.

\section{\bf Nonlinear Gauge, Particles and Fields}

In the preceding section Eq.(56) and Eq.(68) are look-a-like relativistic Hamilton-Jacobi (HJ) equations for a massless particle. In classical mechanics point particle dynamics may be described using the Hamilton principal function $S(q, p, t)$ that assumes the following form if Hamiltonian is a constant of motion
\begin{equation}
 S(q, p, t) = W(q, p) - E t
\end{equation}
Here $W(q, p)$ is the characteristic function, and $(q, p)$ canonical variables. In analogy to optics , $W$ in HJ equation has the same role as the eikonal $L$. Textbook \cite{38}  section (10-8) makes quite insightful remarks on the geometrical optics limit of the light wave and the role of HJ equation in understanding particle trajectory in Schroedinger wave mechanics. The main problem that remains unsatisfactorily resolved till date is the meaning of a localized point particle in the wave description or the continuous field theoretic description.

In the preceding section we have suggested that particle aspect is embodied in a topological defect. Unfortunately the continuous fields of space and time obeying specific partial differential equations seem to have intrinsic limitations for incorporating the topological objects as we have seen for the $\Psi$ field or nonlinear gauge for $A_\mu$ that makes the necessity of the discontinuous field natural, however the particle interpretation and the description of the observed physical phenomena, e. g. electron and EM field interaction and EM waves, in terms of these topological objects are not obvious. Reflecting on his life-long efforts to understand waves and particles de Broglie made profound remarks on them in his essay \cite{39}. Here we note two of them that seem to be useful for the present considerations. First one concerns the definition of a particle as a localized object of high energy concentration that to a first approximation is a shifting singularity. The second point is that the principle of least action is a particular case of the second law of thermodynamics; entropy of a particle is defined by the relation entropy/Boltzmann constant is equal to action/Planck constant. Now thermodynamical concept of a particle is very important: it brings the role of statistical mechanics to the prominence even for an isolated free particle; in de Broglie theory it is caused due to a hidden thermostat. The present work marks a more radical departure than that of de Broglie: space and time are endowed with nontrivial topology, and the geometric objects and fields are proposed to have the statistical nature. 

We elucidate these ideas considering the significance of the nonlinear gauge in the context of \cite{14, 15}. Dirac attempts a modification of CED with the aim to cure the problem of infinities in QED \cite{14} proposing the nonlinear gauge 
\begin{equation}
 A_\mu A^\mu = \frac{m^2 c^4}{e^2}
\end{equation}
The Maxwell action is modified incorporating the constraint (85) using a Lagrange multiplier $\Lambda$. The charge current density is identified to be 
\begin{equation}
 J_\mu = -\Lambda A_\mu
\end{equation}
Dirac argues that if the source-free field has the potentials $A_\mu^*$ then the gauge transformed potentials $A_\mu^* + \Phi_{,\mu}$ corresponds to the charges. Eq.(85) then becomes 
\begin{equation}
 (p_\mu +e A_\mu^*)(p^\mu + e A^{\mu*}) = m^2 c^4
\end{equation}
where $e \Phi{,\mu}$ is interpreted as the energy-momentum 4-vector $p_\mu$ in the relativistic HJ equation (87) for the electron. Note that Dirac develops his theory in flat spacetime. The formal similarity of the charge current density (86) with that of Weyl theory, and for constant $\xi$ with the generalized Weyl-Dirac theory \cite{11} is noteworthy. However in the generalized Weyl-Dirac theory $\xi$ is a co-scalar field, therefore, it cannot be a constant. Moreover the analogue to HJ equation for electron is Eq.(40) that suggests that mass is a space-time dependent field variable, and $\Psi$ field does not appear directly in the current density (42). In contrast to particle aspect introduced through $p_\mu$ in Dirac theory we have the propagating $\Psi$ field discontinuity representing the particle. 

Gubarev et al \cite{15} introduce a novel idea seeking the physical significance of the gauge noninvariant quantity $A_\mu A^\mu$. The volume integral of this quantity in Euclidean space is shown to have minimum value under the gauge transformation for a specific gauge condition. For example, ${\bf B} ={\bf \nabla} \times {\bf A} $ is gauge invariant, and the volume integral $\int {\bf A}. {\bf A} d^3x $ under the gauge transformation is stationary for the Coulomb gauge ${\bf \nabla}. {\bf A} =0$. Authors emphasize the point that their objective is to find the minimum value of this quantity itself that may throw light on the nontrivial topology. This motivation comes from the considerations on the vacuum condensates, for example, quark and gluon condensates in QCD. 

It would, of course, be logically more satisfying to treat $A_\mu$ as a statistical field variable, and interpret the minimum value in terms of the averages 
\begin{equation}
 \sigma (A_\mu) = \overline{A_\mu A^\mu} -\overline{A_\mu}~ \overline{A^\mu}
\end{equation}
The authors \cite{15} do consider the expectation values of the field operators in QFT, however here we associate probability distribution function with the potentials.

Let us recall that the pure gauge potential $\mathcal{A}_\mu$ satisfies the Lorentz gauge condition (20) and the minimum value of the integral of $\mathcal{A}_\mu \mathcal{A}^\mu$ is zero in view of Eq.(66). The potential $\mathcal{A}_\mu$ is harmonic (81) for the nontrivial topology representing a single photon. For a monochromatic photon beam of uniform photon number density it would be natural to assume Eqs. (20) and (66) for the photon beam. In a fluid flow with velocity ${\bf v}$ and probability density $\rho_f$ the local conservation of particles satisfies the continuity equation
\begin{equation}
 \frac{\partial \rho_f}{\partial t} +{\bf \nabla} . \rho_f {\bf v} =0
\end{equation}
Statistical distribution function for photon fluid in analogy to this may be introduced interpreting
\begin{equation}
 {\bf A} = f_p(x_\mu) \mathcal{\bf A}
\end{equation}
as the momentum density and the Lorentz condition as the energy-momentum conservation equation; note that the geometric quantity $\mathcal{A}_\mu$ in Weyl geometry has been multiplied by Planck constant. In \cite{15} the minimum value of the vacuum expectation value of $<0|A_\mu A^\mu|0>$ has been discussed and suggested to be nonzero. In the present paper we have geometry/topology plus statistical/thermodynamical approach: at zero temperature the minimum value is proposed to correspond to the frozen phase of photons with zero momenta and only spinning topological objects possessing energy $\frac{h \nu}{2}$ per photon \cite{12}. It may be contrasted with the assumed ZPF in stochastic electrodynamics, see section 2.2 in \cite{24}.

The main inference that could be drawn from these considerations is that the EM fields in CED have to be interpreted as averaged statistical quantities representing the photon fluid, i. e. rotating fluid of microscopic spinning topological objects, namely the photons.
 Interpreting $A_\mu$ as averaged energy-momentum vector would imply to interpret $F_{\mu\nu}$ in analogy to the antisymmetric second rank angular momentum tensor $L_{\mu\nu}$: ${\bf B}$ as averaged angular momentum (orbital plus spin) and ${\bf E}$ as the averaged value of the energy and the momentum of the fluid as a whole about center of energy; this interpretation is based on the classical relativistic mechanics, see section 14 in \cite{40}.
 
 \section{\bf Discussion and Conclusion}
 
 Radically new approach to geometrize physics is articulated in the present paper: (1) spacetime of relativity has statistical nature generalizing Menger's statistical metric space to Eq.(17), (2) EM potential $\mathcal{A}_\mu$ has fundamental reality representing a single photon as a topological defect in space-time, and (3) gauge conditions in original Weyl geometry, i. e. Weyl natural gauge, Lorentz gauge and nonlinear gauge are proved to be consistent in generalized Weyl-Dirac theory \cite{11} necessitating the existence of topological defects. It is argued that Maxwell field tensor and metric tensor represent statistically averaged quantities.
 
 Space-time itself is visualized as a fluid comprising of microscopic particles defined by topological defects, for example, spinless electron as a moving surface discontinuity and photon as a harmonic 1-form. EM waves would be like sound waves in a fluid whereas photons are localized space-time topological defects, Eq.(81). Interaction of electron with EM field in the present framework is viewed as a random scattering of electron propagating in the photon fluid. Momentum exchange with most probable fraction of photon momentum given by the fine structure constant is envisaged from 
 \begin{equation}
  {\bf p} - \frac{e}{c} {\bf A} ~\rightarrow ~ \frac{\bf p}{\hbar} -\frac{e^2}{\hbar c } {\bf A}_g
 \end{equation}
 where ${\bf A}_g$ is in geometrical units a la Weyl geometry. The most important implication of the present work is that it paves the path for developing an alternative theory of elementary particles and their interactions solely on the geometry and topology of physical space and time \cite{22}. We emphasize that the present ideas are radically new as compared to those of Riemann-Einstein-Weyl on spacetime, however they adhere to the conventional wisdom of space and time reality in contrast to the current radical ideas articulating emergent spacetime and/or discarding space and time reality.
 
 To put the speculations in perspective we present a brief discussion on Einstein's belief \cite{41} and alternative ideas \cite{23, 24, 39}. Einstein in reply to the criticisms \cite{41} states his belief in field theory as a program for physics. Fields are continuous functions in 4D spacetime continuum. Illustrating his idea taking the example of general relativity he puts forward three points. E1: Physical things are described by continuous functions of spacetime, E2: the fields are tensors, e. g. $g_{\mu\nu}$ for gravity, and E3: physical measurability of the invariant line element. According to him the construction of a mathematical theory rests exclusively on E1 and E2. If a complete physics theory exists E3 is not required. Einstein expresses serious reservations on radical efforts of Menger in this perspective. What is the meaning of a mathematical theory of physics? As pointed out in section II mathematical objects like a geometrical point or a circle are mental objects and mathematical logic is sufficient to develop a mathematical realm, for example, Riemannian geometry or Weyl geometry. However physical objects need physicalization or approximation to the mathematical objects, e. g. the fields. Therefore, the geometrization of physics as a program has to alter the Einsteinian ideas, and adopt in some way Clifford and Menger speculations. The present approach has sought this objective based on the Weyl geometry.
 
 The role of probability in fundamental physics distinct than the orthodox quantum theory has also been of interest among some physicists; unfortunately such works have remained sidelined in the mainstream physics literature. A common lament by de Broglie \cite{39} , Luis de la Pena \cite{23} and Boyer \cite{24} is that full potential of their ideas remains unexplored. Could the present modest contribution provide impetus to such efforts? It seems the inclusion of topological model of photon and associated interpretation of zero point energy would be essential to develop a complete stochastic theory for Maxwell equations and Newton-Lorentz equation of motion. 
 
 In conclusion, consistency of various gauge conditions in Weyl geometry is thoroughly studied and the physical reality of EM potentials is established. Harmonic 1-form in 4D spacetime of Weyl needs further investigation in connection with the physical mechanism of the AB effect \cite{6}. Novel aspects on the topology and statistical nature of space and time geometry proposed here would strengthen the outlook on particle physics based on the real wave equations in \cite{22}.

 {\bf APPENDIX-I}

Levi-Civita and Weyl define parallel displacement of a vector $V_\mu$ from spacetime point $x^\mu$ to $x^\mu +\delta x^\mu$; the change of the vector round a small loop is given by
\begin{equation}
 \delta V_\mu =\frac{1}{2} (V_{\mu:\nu:\sigma} - V_{\mu:\sigma:\nu}) ~dS^{\nu\sigma}
\end{equation}
Here  $V_{\mu:\nu:\sigma}$ is second covariant derivative of $V_\mu$. In the Riemannian space, using the definition of the Riemann curvature tensor we have
\begin{equation}
 V^\mu \delta V_\mu = \frac{1}{2} R_{\mu\nu\sigma\lambda} V^\mu V^\lambda ~dS^{\nu\sigma} =0
\end{equation}
since $R_{\mu\nu\sigma\lambda}$ is antisymmetrical in $\mu$ and $\lambda$. Thus the length of the vector $l=V^\mu V_\mu$ does not change under parallel transport round the loop in the Riemannian space.

In Weyl space, the generalized Christoffel symbol (25) shows that the curvature tensor $^*R_{\mu\nu\sigma\lambda}$ has a part symmetrical in $\mu$ and $\lambda$ \cite{7, 13}. This part gives the nonintegrability of the length
\begin{equation}
 \oint \delta l =2 V^\mu \delta V_\mu = W_{\nu\sigma}~ g_{\mu\lambda} V^\mu V^\lambda ~dS^{\nu\sigma}
\end{equation}

{\bf APPENDIX-II}

Here we derive Eq.(41) using the variational principle $\delta S=0$ from Eq.(39) considering the variations in $A_\mu$ treated as independent field variable. We get the following variations
\begin{equation}
 p \delta(F^{\mu\nu} F_{\mu\nu} \sqrt{-g}) = -4 p F^{\mu\nu}_{~:\nu} \sqrt{-g} ~ \delta A_\mu
\end{equation}
\begin{equation}
 \xi \delta(A^\alpha_{~:\alpha} \sqrt{-g}) = - \xi^{:\mu} \sqrt{-g} ~\delta A_\mu
\end{equation}
\begin{equation}
 \xi \delta (A^\alpha A_\alpha \sqrt{-g}) =2 \xi A^\mu \sqrt{-g}~ \delta A_\mu
\end{equation}
Substituting expression (16) in Eq.(39), and using (95)-(97) we finally arrive at the modified Maxwell field equation (41).

Apparently current density (31) in Weyl theory; (41) in generalized Weyl-Dirac theory; and (86) in new electron theory of Dirac seem to be in conflict with the standard CED. On closer examination it is found that there are physical arguments showing that there is no contradiction. Recall that in the Maxwell-Lorentz theory the charge current density four-vector originates from the macroscopic system of charges and currents, however for a superconductor London assumed ${\bf J}$ proportional to ${\bf A}$. In quantum theory the probability current density for a Schroedinger charged particle interacting with the magnetic field acquires a component proportional to the vector potential ${\bf A}$. For a Dirac electron in the presence of the EM field the Dirac current $\bar{\Psi} \gamma^\mu \Psi$ is unaltered, however performing Gordon decomposition it is found that the Gordon current is modified containing a term proportional to the EM potential $A_\mu$. For further details we refer to \cite{42}.

{\bf ACKNOWLEDGMENT}

I gratefully acknowledge the positive outlook and useful comments of the reviewers.

\end{document}